\documentclass[usenatbib,useAMS]{eas}
\usepackage{pslatex,graphicx}
\usepackage{amsmath,amsfonts}
\usepackage{pdfscreen}

\def\idm#1{{\mbox{\scriptsize #1}}}

\begin{document}

\title{Modeling the RV and BVS of active stars} 
\runningtitle{RV and BVS analysis}
\author{Cezary Migaszewski}
\address{Toru\'n Centre for Astronomy, Nicolaus Copernicus University,
Gagarin Str. 11, 87-100 Toru\'n, Poland;
\email{[c.migaszewski,g.nowak]@astri.umk.pl}}
\author{Grzegorz Nowak}
\sameaddress{1}
\begin{abstract}
We present a method of modeling the radial velocity (RV) measurements which can
be useful in searching for planets hosted by chromospherically active stars. We
assume that  the observed RV signal is induced by the reflex motion of a star as
well as by distortions of spectral line profiles, measured by the Bisector
Velocity Span (BVS). The RVs are fitted with a common planetary model including
RV  correction term depending linearly on the BVS, which accounts for the
stellar activity.  The coefficient of correlation is an additional free
parameter of the RV model.  That approach differs from correcting the RVs {\em
before} or {\em after} fitting the ``pure'' planetary model.  We test the method
on simulated data derived for single-planet systems. The results are compared
with the outcomes of algorithms found in the literature. 
\end{abstract}
\maketitle
%
\section{Introduction}
The current spectral Doppler technique makes it possible to measure routinely 
the radial velocity (RV) of stars with a precision better than
$10~\mbox{ms}^{-1}$.  It is well known that the RV variability may be caused not
only by the reflex motion of a star accompanied by smaller bodies, but also by
internal activity of stellar atmosphere (e.g., non-radial pulsations,
inhomogeneous convection, and/or rotating surface spots).  While the reflex
motion of the star leads to the Doppler shift of spectral lines, the atmospheric
activity may cause distortions of spectral lines profiles (SLP), which displace
their measured minima. Fortunately, a determination of the true origin of the
RV  variability is possible through the analysis of line profiles or the
cross-correlation function (CCF). The basic tool for such analysis is the line
bisector (LB) technique (\cite{Gray1983,Gray2005}).   The most simple and useful
measure of the slope of LB is the Bisector Velocity Span (BVS) defined as the
difference between the LB velocity measured at some upper and lower flux levels
of the SLP.

It is well known (\cite{Desort2007}, and references therein), that distortions
of SLP caused by  stellar activity may produce  quasi-periodic RV signals
mimicking planetary companions. The BVS analysis are then very helpful to reject
or confirm the planetary nature of the RV variability. The first case, when the
BVS measurements were used to withdraw a hypothesis of a low-mass planet, was 
HD~166435 (\cite{Queloz2001}). In the next years, other stars with low-level RV
variations stemming from stellar activity were discovered, for instance, 
HD~78647 (\cite{Setiawan2004}), and  HD~219542B (\cite{Desidera2004}). In the
first two instances, the authors relied on a correlation between the RV and the
BVS data, indicating that periodicity of the RV signal has the stellar-activity
origin. (Further in this paper  we will demonstrate  that, in general, such a
correlation does not provide sufficient information to reject the planetary
hypothesis).  Recently, \cite{Bonfils2007} report a discovery of  a
``super-Earth'' planet orbiting GJ~674 with a 4.69-day period. Besides the
strong signal of that planetary candidate,  the RV data reveal also secondary,
34.85-day periodicity. The discovery team performed 2-planet Keplerian fits and,
using photometric measurements,  attributed the second period to the rotational
modulation  of the RV caused  by a stellar spot.

In this work, we propose an alternative method of resolving non-unique sources
of the RV variability. It relies on simultaneous analysis of both sets of
observables, RV and BVS, within an uniform fit model.  In Section~2, we
introduce the generalized model of RV. In Section~3, we simulate RV data for our
experiments. Section 4 is devoted to  RV models and the results, which are
finally discussed in  Conclusions.

%
\section{Modeling the RV and BVS data}
%
The stellar RV variability caused by the presence of additional bodies may be
modeled as a superposition of Keplerian, astrocentric orbits 
(\cite{Smart1949}):
\begin{equation}
V_r(t) = \sum_{i=1}^N {K_i \big[\cos{(\varpi_i+\nu_i)} + e_i
\cos{\varpi_i}\big]} + \sum_{o=1}^O{V_o},
\label{model1}
\end{equation}
where $N$ is the number of planets, {$K_i$} is the semi-amplitude of the $V_r$
contribution by the {$i$}-th planet, {$\varpi_i$} is for the longitude of
pericenter, {$e_i$} is for the eccentricity, {$\nu_i$} denotes the true anomaly
(which depends implicitly on the orbital period {$P_i$} and the time of
pericenter passage {$\tau_i$}), {$V_o$} are the RV offsets and $O$ is their
number. We search for such parameters  {$(K_i, P_i, e_i, \varpi_i, \tau_i,
V_o)$,  which may explain the RV variability in the sense of the least squares.

When we deal with significant stellar activity, and the BVS measurements are
available, we can modify the RV model by adding a correction term, accounting
for distortions of the SLP which contribute to the RV variability, 
$\mathcal{V}_r(t) = V_r(t) + \Delta V(t)$.  In the  first approximation,
{$\Delta V(t)$}  may be expressed through the BVS time series ($\{BVS\}$ from
hereafter), $\Delta V(t_j) = \alpha \{BVS\}_j, (j=1,\ldots,N_m),$  where $N_m$
is the number of observations and  $\alpha$ is {\em a free parameter} of the
model. In fact, $\alpha$  depends not only on stellar activity but also on the
spectral resolution (\cite{Desort2007}),  and a particular choice of the upper
and lower segments of the LB.  Now, following the principle of the least
squares, we define the $(\chi_r^2)^{1/2}$ function of the fit model as follows:
\begin{equation}
\left(\chi_r^2\right)^{1/2} = \left(\frac{1}{N_{m} - N_{p} - 1}
\sum_{j=1}^{N_{m}} \frac{\big[ \{RV\}_j - \alpha \{BVS\}_j - V_r(t_j) \big]^2}
{\sigma_{{RV},j}^2 + \alpha^2
\sigma_{\textrm{BVS},j}^2}\right)^{1/2}, 
\end{equation}
where {$N_{p}$} is the number of model parameters, {$\sigma_{{RV},j}$}, 
{$\sigma_{\textrm{BVS},j}$} stand for the standard errors of the $\{RV\}$ and
$\{BVS\}$ time series,  respectively. 

It is worth to note that the BVS may be used to correct the RVs for  stellar
activity contribution not only when the central star exhibits periodic variability,
but also when that activity has an irregular (aperiodic) character.  In fact,
the BVSs measure shifts of detected minima of spectral lines, regardless of 
a type and sources of stellar activity. 
%
\section{Simulated observations of the RV and BVS}
%
To give an example, and to demonstrate features of our method, we test model
Eq.~2.2 on synthetic RV and BVS observations. We consider a star exhibiting 
periodic activity that mimics planetary RV signal, and we assume that it also
hosts a planet. To construct the synthetic data set, we fix the elements of
Keplerian orbit, and we simulate the planetary RV signal, 
$\{RV\}^{\idm{(pl)}}$, Eq.~\ref{model1}. We also simulate the $\{BVS\}$ data.
The synthetic signal,  $\{RV\} = \{RV\}^{\idm{(pl)}} + {\alpha} \{BVS\}$. 
Next,  we add Gaussian errors independently to both sets of observables and also
non-periodic noise (jitter) which is added to  the $\{RV\}$ and $\{BVS\}$ in the
same phase. Parameters of the synthetic signals are given in caption to
Fig.~\ref{data}. 

For a reference, we assume that a single-planet system  has the same orbital and
stellar-activity periods. Our choice of stellar parameters follows the first
discovery of a star that mimics the  planetary signature of RV, i.e., HD~166435
(\cite{Queloz2001}). The synthetic curves of the RV and BVS, as well as
simulated measurements (the red points) in random epochs, are illustrated in
Fig.~\ref{data}. The first two panels are for the $\{RV\}$  and $\{BVS\}$ time
series. The third panel is for a correlation between these observables. Grey,
thick line illustrates the linear regression of the stellar-activity
contribution to the RV ($\{RV\}^{\idm{(st)}}$) on $\{BVS\}$. Black, thick line
marks the  linear correlation between the $\{RV\}$ and $\{BVS\}$ sets.
\begin{figure*}
\centerline{
\includegraphics[width=12.5cm]{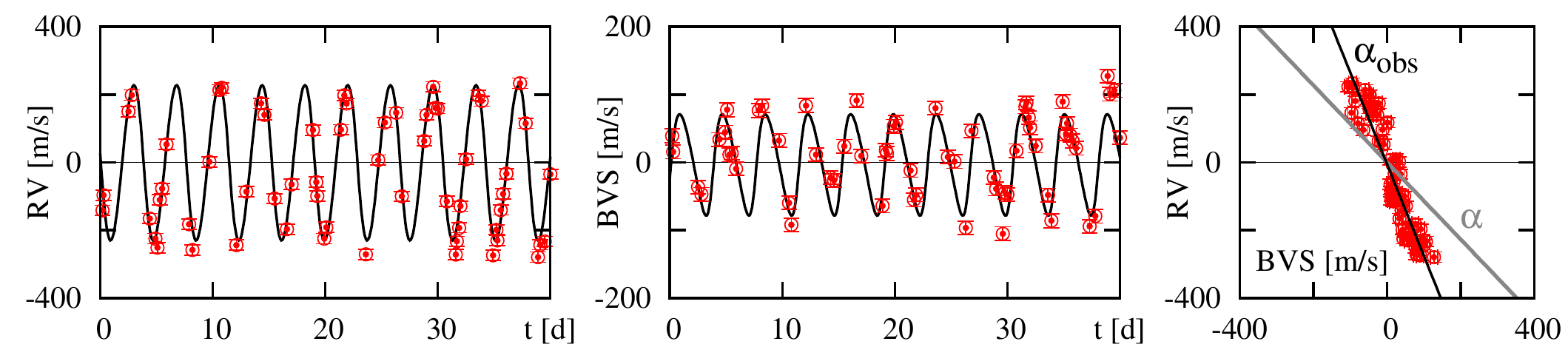}
}
\caption{
Simulated $\{RV\}$ and $\{BVS\}$ data for the following parameter  tuples of $(K
[\mbox{ms}^{-1}], P [\mbox{d}], e, \varpi [^o],  \mathcal{M}_0 [^o])$;  {\em
stellar-activity signal:} $(85, 3.8, 0.2, 75, 0)$;  {\em planetary signal:}
$(150, 3.8, 0, 270, 180)$; stellar jitter is  $30~\mbox{ms}^{-1}$, $\sigma_{RV}
= 15~\mbox{ms}^{-1}$,  $\sigma_{BVS} = 10~\mbox{ms}^{-1}$, $\alpha = -1.1364$.
}
\label{data}
\end{figure*}
%
\section{Tests of fitting algorithms and the results}
The results of analysis of the synthetic data are illustrated in
Fig.~\ref{test1}.  Panels in this figure are for color-coded maps of
{$(\chi_r^2)^{1/2}$} in  planes of selected fit parameters.  Contours mark the
standard confidence intervals of  ({$1\sigma, 2\sigma, 3\sigma$}), respectively.
The filled, crossed circles mark positions of the nominal  solution in the
{$(P_b, e_b)$}-- and {$(P_b, K_b)$}--planes (the left- and the right-hand
columns, respectively). To compute the $(\chi_r^2)^{1/2}$-maps, we proceed as
follows.  We fix a particular  point $(x,y)$ in the selected parameter plane, 
and we search for remaining best-fit elements of the RV model. To search for the
best-fit solution,  we apply the Monte-Carlo method to choose  initial
conditions for the fast Levenberg-Marquard algorithm (see, e.g.,
\cite{Gozdziewski2008}). Such fits are repeated for all points $(x,y)$ of a
discrete grid in the given plane of model parameters.
\begin{figure*}
\centerline{
\hbox{\includegraphics[width=49mm]{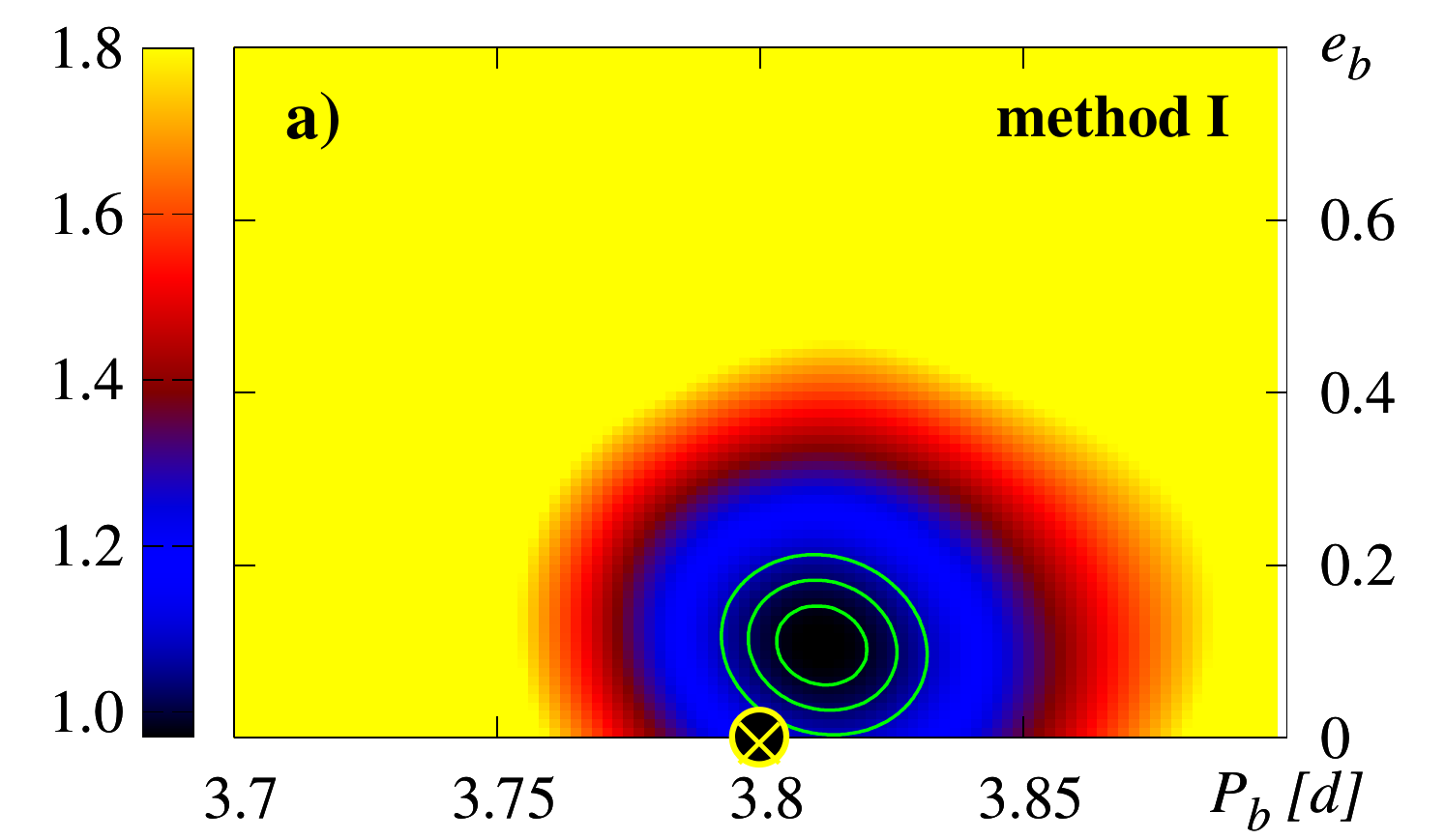}\hskip8mm
      \includegraphics[width=49mm]{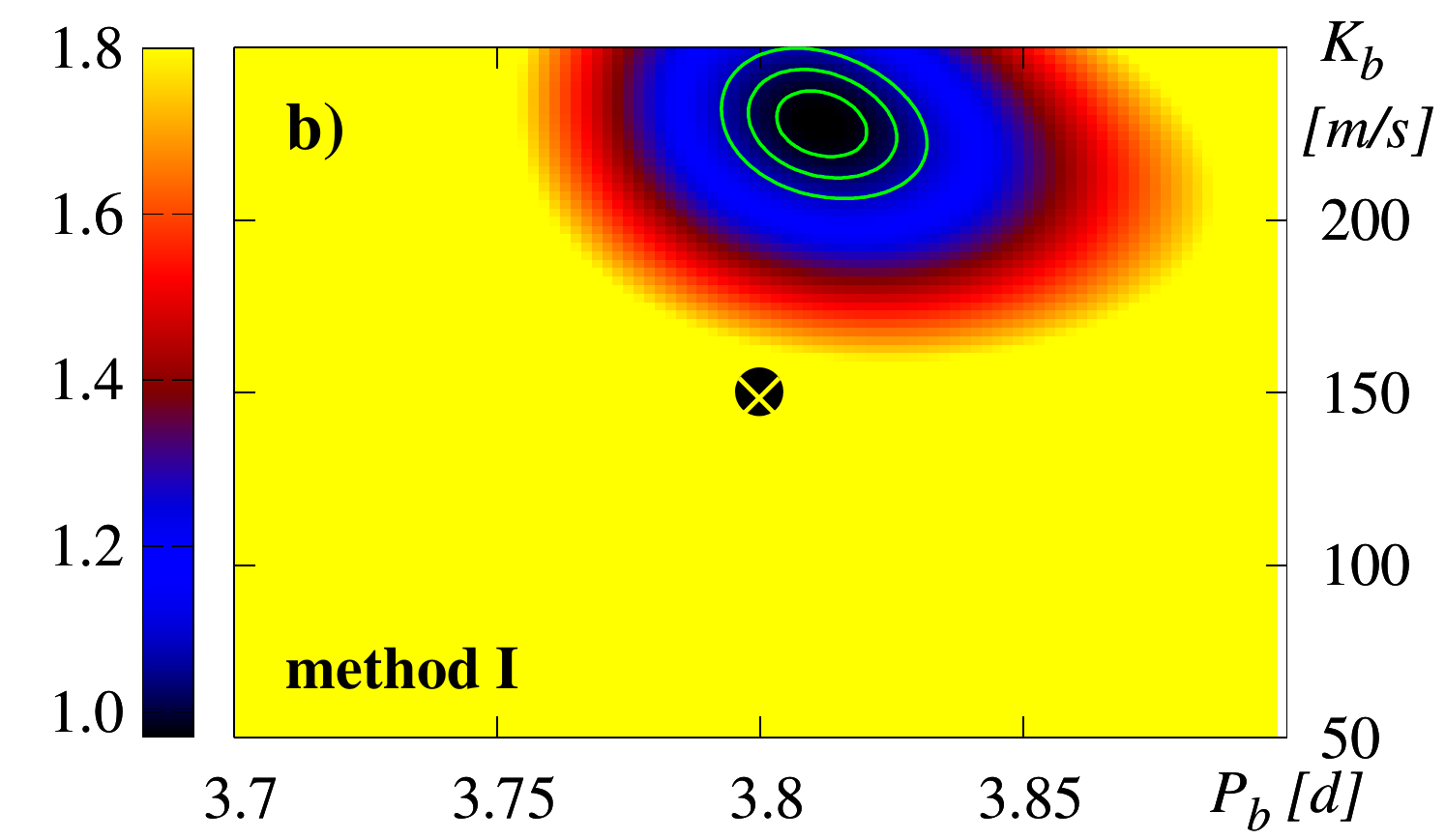}
     }
}
\centerline{
\hbox{\includegraphics[width=49mm]{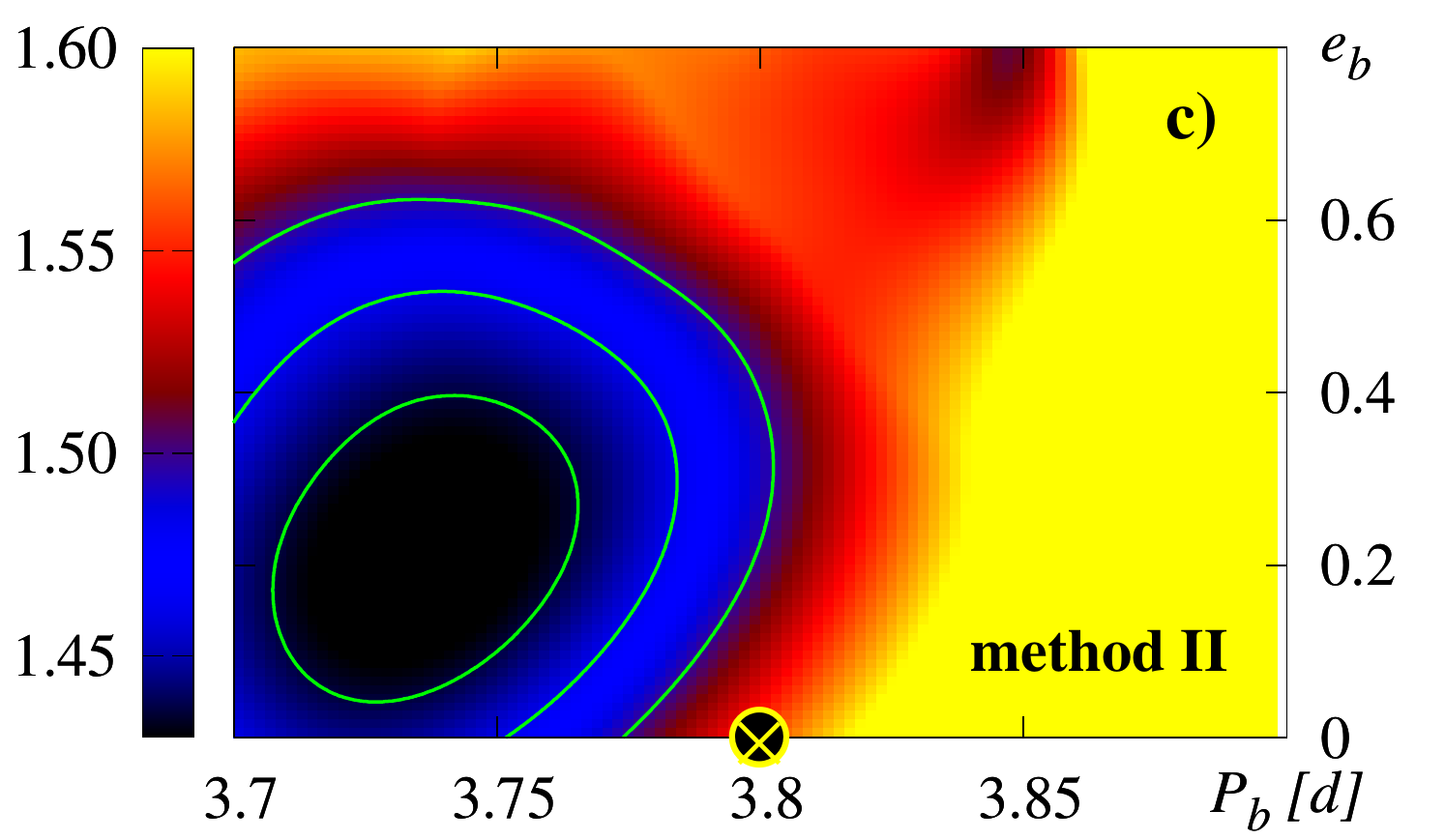}\hskip8mm
      \includegraphics[width=49mm]{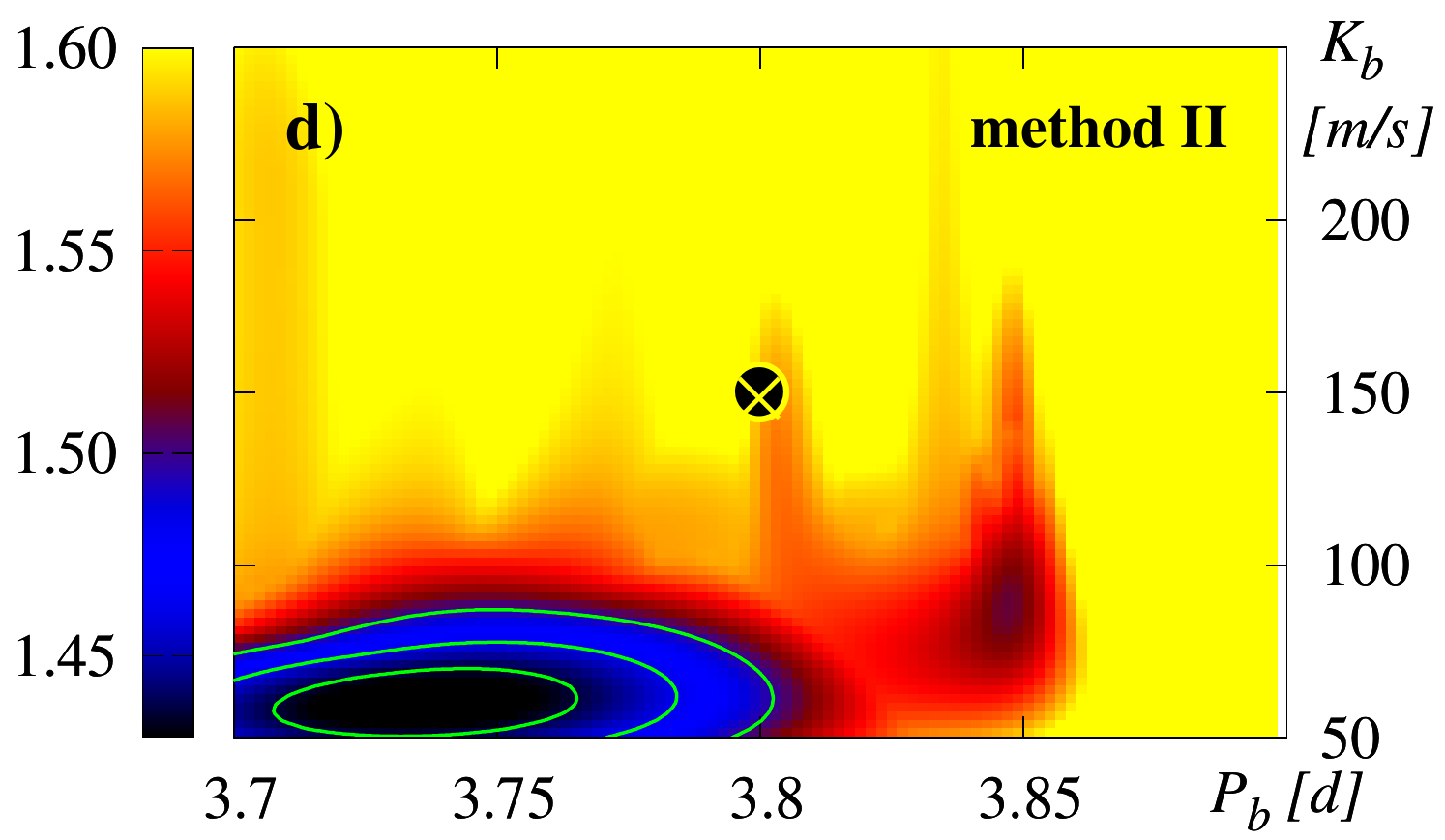}
     }
}
\centerline{
\hbox{\includegraphics[width=49mm]{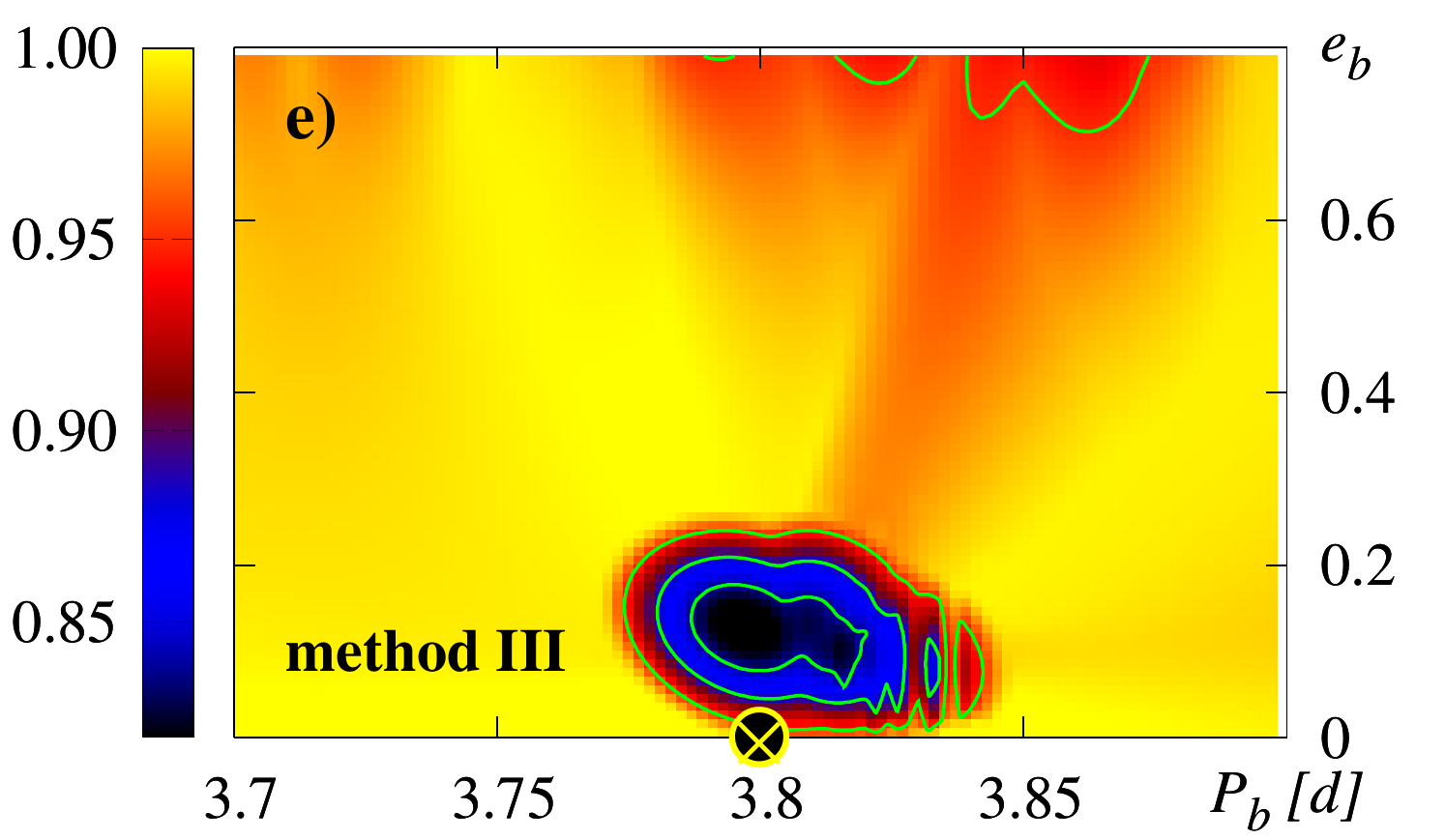}\hskip8mm
      \includegraphics[width=49mm]{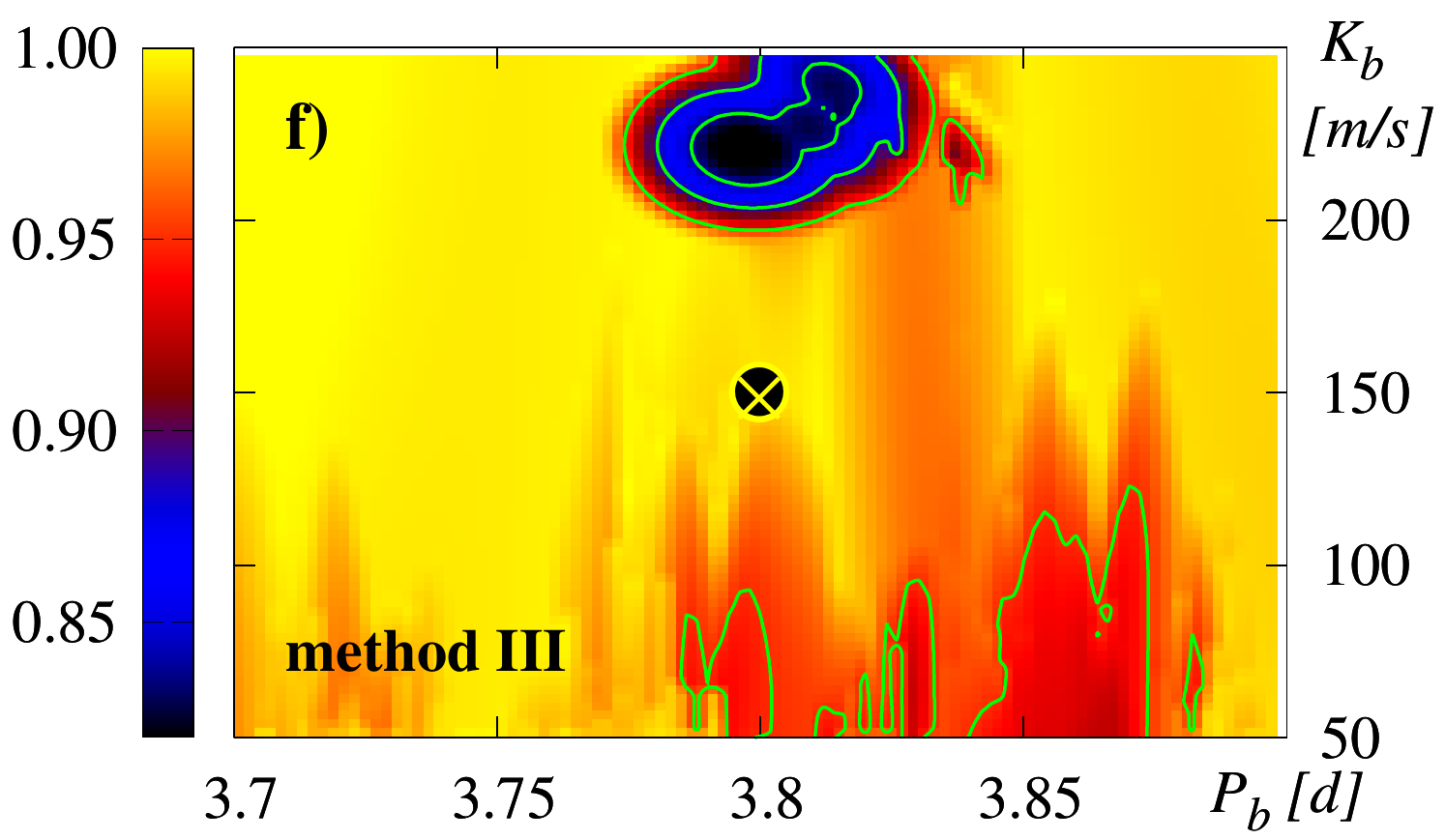}
     }
}
\centerline{
\hbox{\includegraphics[width=49mm]{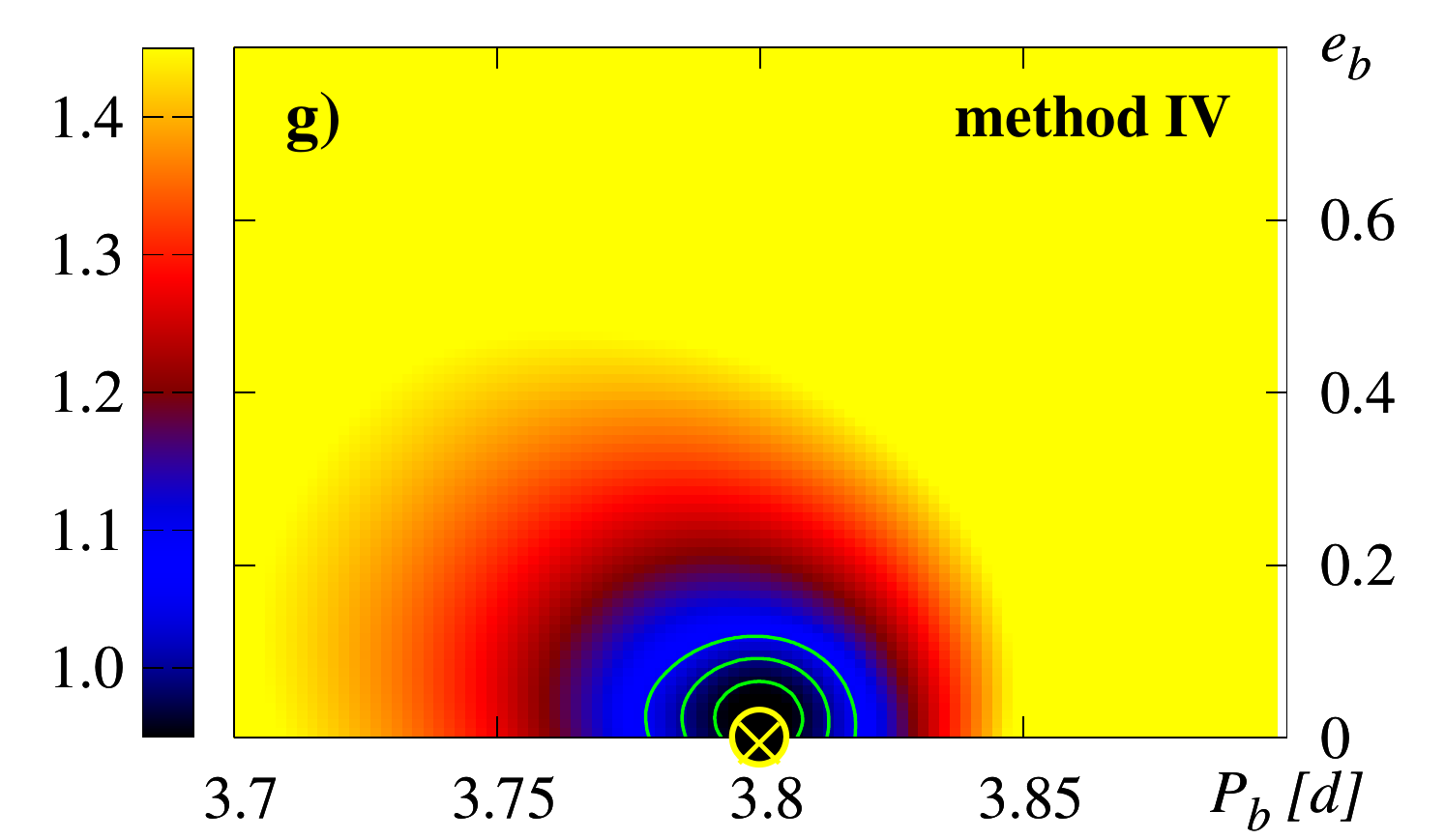}\hskip8mm
      \includegraphics[width=49mm]{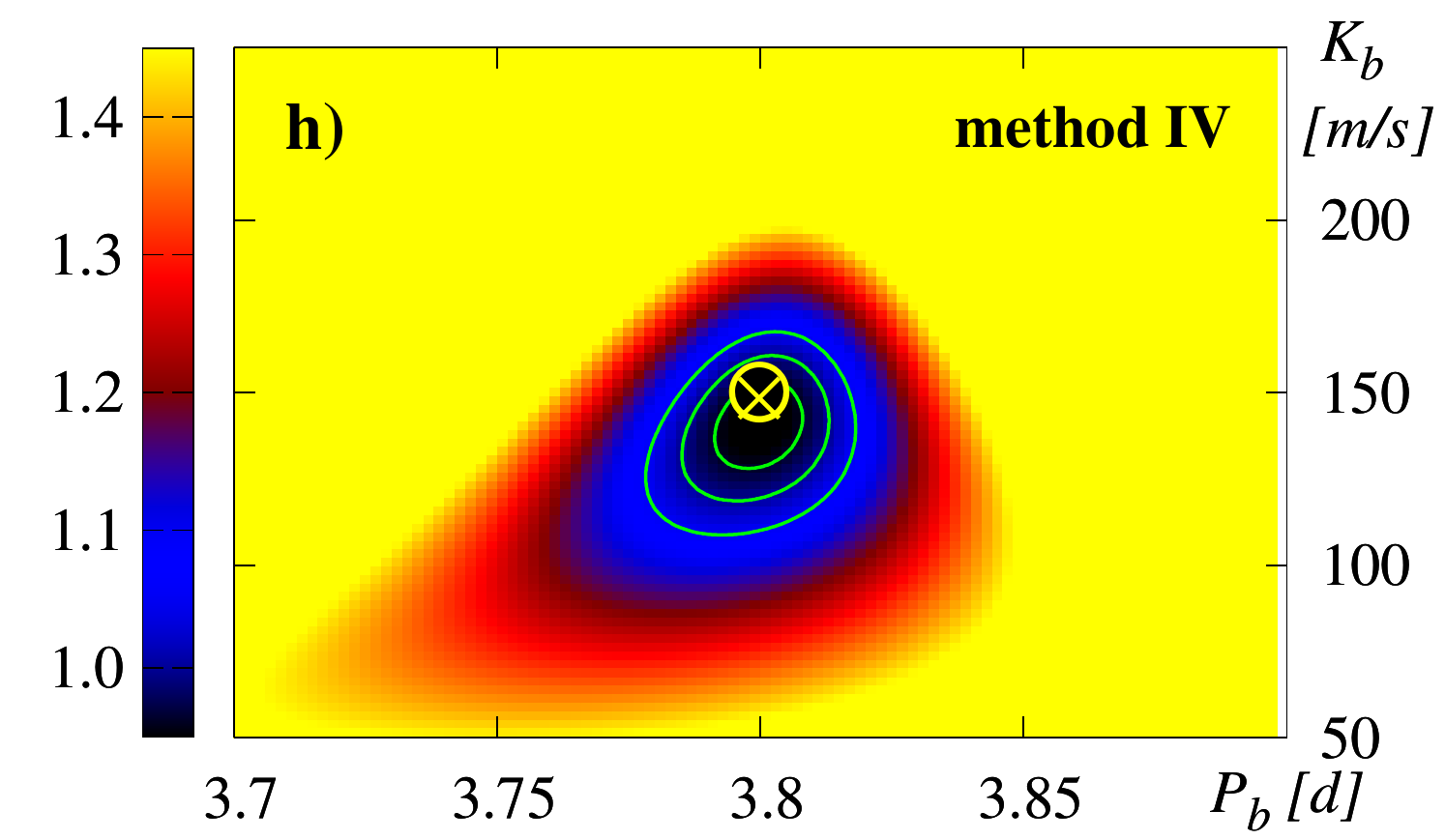}
     }
}
\caption{ 
Maps of $(\chi_r^2)^{1/2}$ for the test case. Simulated data are presented in
Fig.~\ref{data}.  {\em The left column}: the $(P_b, e_b)$-plane;  {\em the right
column}: the $(P_b, K_b)$-plane. Each row is for a different fit algorithm
tested in this paper (see the text for details). Filled, crossed circles mark
the orbital parameters of the nominal system (given in caption to
Fig.~\ref{data}). 
}
\label{test1}
\end{figure*}

Each row in Fig.~\ref{test1} is for a different method of modeling the RVs.  We
start from the most simple (but generally wrong) method~I that relies on fitting
single-planet model (Eq.~\ref{model1}) to the $\{RV\}$ signal, without making
any use of  additional information ``hidden'' in the BVS. The results are
illustrated in Fig.~\ref{test1}a,b. Clearly, the nominal solution lies beyond
the {$3\sigma$} level of the best fit solution. In particular, the fitted
{$K_b$} is significantly larger than the reference value.  Both signals,
$\{RV\}^{\idm{(pl)}}$ and $\alpha~\{BVS\}$, have the same periods, and depending
on their  relative phase, the semi-amplitude of the resulting $\{RV\}$ may be
larger or smaller than  that ones of the planetary signal. In the tested
example, the $\{RV\}$ signal has larger amplitude than the
$\{RV\}^{\idm{(pl)}}$. Hence, we obtain  {\em too large}  mass of the planetary
companion, however, the original orbital period is found correctly.

To improve method~I, one may proceed as follows (method~II). At first, we find 
the coefficient $\alpha_{\idm{obs}}$ of the linear correlation  between the
$\{RV\}$ and $\{BVS\}$ signals. Then we subtract a correction term, 
$\alpha_{\idm{obs}}\{BVS\}$, from the observed $\{RV\}$ signal. We obtain, as we
suppose, a pure planetary signal, so we can fit the single-planet model to
corrected $\{RV\}$ data. The results are presented in Fig.~\ref{test1}c,d.
Again, the derived best-fit solution is displaced from the true position (in
particular, its semi-amplitude is badly determined). To explain that improper
outcome of the fit,  we look at the right-hand panel of Fig.~\ref{data}. The
grey line is for the correlation between $\{BVS\}$ and $\{RV\}$ observables when
the star would be alone. If the planet is present then the detected correlation
($\alpha_{\idm{obs}}$, black line) changes. The difference between $\alpha$ and
$\alpha_{\idm{obs}}$ depends on the semi-amplitudes of both signals as well as
on their  relative phase. Unfortunately, also  uncertainties of the
measurements  (instrumental ones or stemming from the stellar jitter), as well
as their irregular sampling may change the correlation coefficient. Then, if we
use $\alpha_{\idm{obs}}$ to correct the RVs by the linear term of
$\alpha_{\idm{obs}}\{BVS\}$, we may subtract too large or too small correction
from the $\{RV\}$ signal, and the best fit parameters of the orbital solution 
will be wrong.  In contrary to method~I, we obtain {\em too small} mass of the
planet.

The next approach relies on fitting 2-planet Keplerian model to the $\{RV\}$
data (method III, see \cite{Bonfils2007}). We obtain two best-fit Keplerian
orbits, moreover, we have to identify which orbit describes the true reflex
motion of the star, and which one corresponds to a false orbit mimicked by 
stellar activity. We did a test of this method and the results are illustrated
in Fig.~\ref{test1}e,f.  Still, we  do not obtain correct results.  The main
difficulty in this approach emerges from a non-unique identification of the
signals. For instance, a sum of two different (de-phased) RV signals with the
same period may look very similar to a composition of two signals with the $2:1$
commensurability  of periods (see, e.g., \cite{Gozdziewski2006}).  In fact,  the
second orbital period found in the best fit solution,  may be very different
from its real value.  Method~III gives also too large mass of the planet.

Finally, we apply our method~IV relying on simultaneous analysis of both sets
of  $\{RV\}$-- and $\{BVS\}$--observables. We assume that {$\alpha$}  is an
additional, free parameter of the fit model. Actually, we found that this
approach provides the best results, as compared to the outcomes of
methods~I--III. That is illustrated in Fig.~\ref{test1}g,h. In this case, the
agreement of the reference parameters with the orbital elements of the best-fit
solution is basically perfect. Thanks to the self-consistent fitting process, we
also find the true correlation coefficient $\alpha$ of the stellar contribution
to the RV, $\{RV\}^{\idm{(st)}}$, and the BVS.
%
\section{Conclusions}
%
Recently, the line bisectors are routinely derived from the same spectra used to
measure the RVs (e.g., the I$_2$-cell technique, \cite{Martinez2005}). They are
much more useful to understand the origin of the RV variations than alternative,
even more indirect  observables.  The BVS data may help to detect
stellar-induced periodicity in the RV signals or to correct the RVs by
accounting for the linear correlation between the RV and BVS data.  In this
work, we also show that these data make it possible to find correct orbital
parameters through fitting the planetary model to the RV and BVS data
simultaneously. Our method seems to work well, even if there is a close
commensurability between periods of both signals. In such a case,  other
algorithms tested here, provide the best-fit orbital elements which are
significantly different from the true, reference solution.

We note that the proposed method omits one more indicator of a distortion of the
SLP, the Bisector Curvature (BC).  The BCs measure the second order derivative
of the LB. It still may be possible to improve the algorithm by using the BCs to
correct $\Delta V(t)$. We are going to investigate such an improved method with
more sophisticated and realistic simulations of the measurements. 
%
\section*{Acknowledgments}
%
Many thanks to Krzysztof Go{\'z}dziewski for a discussion and corrections of the
manuscript. This work is supported by the Polish Ministry of Sciences and Higher
Education, Grant No. 1P03D-021-29 (CM), Grant No. 1P03D-007-30 (GN) and by the
Nicolaus Copernicus University Grant No. 408A (CM). GN is a recipient of a
graduate stipend of the Chairman of the Polish Academy of Sciences.

\bibliographystyle{astron}
\bibliography{ms}
\label{lastpage}
\end{document}